\documentclass[preprint,review,12pt]{elsarticle}

\usepackage{amssymb}
\usepackage{amsmath}

\usepackage{bm}
\usepackage{mathtools}
\usepackage{caption}
\usepackage{graphicx}
\usepackage{latexsym}
\usepackage{booktabs}
\usepackage{multirow}
\usepackage{algorithm}
\usepackage{algpseudocode}
\usepackage{physics}
\usepackage{xcolor}
\usepackage{pgf}
\usepackage{braket}
\usepackage{autobreak}
\usepackage{makecell}
\usepackage{verbatim}
\usepackage{geometry}
\usepackage{subcaption}
\usepackage{hyperref}
\usepackage{booktabs}
\usepackage{siunitx}
\usepackage{float}
\usepackage{setspace}
\usepackage{tcolorbox}
\tcbuselibrary{listings}
\tcbuselibrary{breakable}
\tcbset{
    listing engine=listings,
    colback=white!95!blue,
    colframe=blue!75!black,
    listing options={
        basicstyle=\ttfamily,
        numbers=left,
        numberstyle=\tiny,
        stepnumber=1,
        showstringspaces=false,
        breaklines=true,
        breakatwhitespace=true,
        tabsize=4,
        language=C++
    }
}

\usepackage{listings}
\usepackage{xcolor}

\lstdefinestyle{python_style}{
    language=Python,
    basicstyle=\ttfamily\small, 
    keywordstyle=\color{blue},  
    commentstyle=\color{gray},  
    stringstyle=\color{olive},  
    numbers=none,              
    frame=none,                  
    tabsize=4,                 
    showstringspaces=false     
}

\lstdefinestyle{text_style}{
    basicstyle=\tt,
    frame=none,
    breaklines=true,
    columns=fullflexible,
    backgroundcolor=\color{gray!20},
    framesep=5pt,
    xleftmargin=0pt,
    xrightmargin=0pt,
    breakindent=0pt
}

\graphicspath{{Fig/}}
\graphicspath{{Fig/template/}}

\usepackage{lineno}

\journal{}

\begin{document}

\begin{frontmatter}



\title{CFD-copilot: leveraging domain-adapted large language model and model context protocol to enhance simulation automation}



\author[fir]{Zhehao Dong}

\author[fir]{Shanghai Du}

\author[fir]{Zhen Lu\corref{cor1}}\ead{zhen.lu@pku.edu.cn}

\author[fir,sec]{Yue Yang\corref{cor1}}\ead{yyg@pku.edu.cn}

\cortext[cor1]{Corresponding author.}

\address[fir]{State Key Laboratory for Turbulence and Complex Systems, School of Mechanics and Engineering Science, Peking University, Beijing 100871, China}
\address[sec]{HEDPS-CAPT, Peking University, Beijing 100871, China}

\begin{abstract}
Configuring computational fluid dynamics (CFD) simulations requires significant expertise in physics modeling and numerical methods, posing a barrier to non-specialists.
Although automating scientific tasks with large language models (LLMs) has attracted attention, applying them to the complete, end-to-end CFD workflow remains a challenge due to its stringent domain-specific requirements. 
We introduce CFD-copilot, a domain-specialized LLM framework designed to facilitate natural language-driven CFD simulation from setup to post-processing. 
The framework employs a fine-tuned LLM to directly translate user descriptions into executable CFD setups.
A multi-agent system integrates the LLM with simulation execution, automatic error correction, and result analysis. 
For post-processing, the framework utilizes the model context protocol (MCP), an open standard that decouples LLM reasoning from external tool execution. 
This modular design allows the LLM to interact with numerous specialized post-processing functions through a unified and scalable interface, improving the automation of data extraction and analysis.
The framework was evaluated on benchmarks including the NACA~0012 airfoil and the three-element 30P-30N airfoil. 
The results indicate that domain-specific adaptation and the incorporation of the MCP jointly enhance the reliability and efficiency of LLM-driven engineering workflows.
\end{abstract}


\end{frontmatter}



\section{Introduction}
\label{sec:introduction}

Computational fluid dynamics (CFD) is a cornerstone of modern aerospace engineering~\cite{spalart2016role,chen2013constrained,aliabadi1995parallel,QUAN2025104227,Ren2014Numerical,Yang2023Applications}, serving as an indispensable tool for design and analysis.
However, the utilization of CFD is hindered by its steep learning curve.
Effectively conducting a simulation requires users to possess deep, interdisciplinary knowledge spanning fluid mechanics, numerical methods, and software-specific configurations. 
This high expertise barrier limits its accessibility, preventing many engineers and researchers from leveraging its full potential.

The recent development of machine learning has led to increasing interest in its application to problems in fluid mechanics~\cite{Brunton2024Promising,Vinuesa2022Enhancing,Zhang2025Physicsinformed,xinlong2025predicting,yao2025rapidly,xiong2023point,yin2020feature,runze2023transfer,zhu2019machine,zhu2021turbulence,zhou2021wall,zhou2022robust,Hu2024Improving,Hu2025Generative}. 
Among these developments, large language models (LLMs)~\cite{guo2025deepseek,achiam2023gpt,yang2025qwen3technicalreport,devlin2019bert} have attracted global attention.
Built upon the transformer architecture and trained on internet-scale corpora, these models have demonstrated powerful capabilities in complex reasoning, natural language understanding, and code generation.
Furthermore, advancements such as prompt engineering~\cite{sahoo2025promptengineering}, retrieval-augmented generation~\cite{lewis2020retrieval} and multi-agent systems~\cite{han2025llmmultiagentsystems} are transforming LLMs from simple text generators into intelligent agents. 
These agents can orchestrate entire simulation workflows by interfacing with external tools, executing code, and analyzing outputs~\cite{bi2024oceangpt,ding2025scitoolagent,luo2025oneke}.
By leveraging these capabilities, LLMs have the potential to automate complex tasks from configuration file generation to post-processing, fundamentally lowering the expertise barrier and making advanced simulation technologies more accessible.

Recent studies~\cite{Xu2024LLM,Chen2024MetaOpenFOAM,Chen2025MetaOpenFOAM,chen2025optmetaopenfoam,Pandey2025OpenFOAMGPT,Feng2025OpenFOAMGPT,Xu2025Engineeringai,Wang2025status,fan2025chatcfd,Wang2025Evaluations,Xu2025CFDAgent,yue2025foamagent,dong2025autocfd,Zhang2025Envision} have explored the integration of LLMs into CFD workflows, particularly through automating open-source platforms such as OpenFOAM~\cite{jasak2007openfoam}. 
While these pioneering works demonstrate the feasibility of using LLMs for simulation setup and post-processing, they also reveal key limitations that hinder their widespread adoption. 
First, many existing approaches rely on knowledge bases constructed from a limited set of predefined tutorials~\cite{chen2025optmetaopenfoam}. This dependence restricts their ability to generalize to novel or non-standard simulation scenarios that deviate from the training examples. 
Second, the absence of a modular function-calling mechanism results in rigid coupling to specific post-processing utilities. These constraints hinder scalability, adaptability, and long-term maintainability, preventing current frameworks from becoming robust, general-purpose solutions for CFD automation.
Consequently, current work lacks a framework that can offer more generalized reasoning, while supporting a more modular and scalable method for tool integration in CFD automation.

The integration of LLMs with external tools has advanced significantly with the development of function calling~\cite{kim2024anllmcompiler}, which has enabled LLMs to invoke external application programming interfaces in a structured manner. 
This capability extends the functionality of LLMs beyond text generation, allowing them to act as agents that can retrieve real-time data, execute domain-specific operations, and interact with software systems~\cite{erdogan2024tinyagent}.
However, conventional implementations of function calling present challenges for scalability and maintainability. Adding a new tool typically requires the manual definition of a corresponding function and schema~\cite{liu2025toolace}.
In addition, many function definitions are tied to a single LLM provider, which means that changing to a different model often requires rewriting the integration.
For domains like CFD post-processing, where many tools are needed and workflows keep changing, this approach is difficult to maintain.

The model context protocol (MCP)~\cite{anthropic2025mcp} addresses these limitations by offering an open standard for tool integration. 
It decouples the reasoning capabilities of LLMs from the execution of external tools through a standardized communication interface~\cite{hou2025modelcontextprotocolmcp}. 
Just as USB-C provides a common standard for hardware devices, MCP allows tools and data sources to be built once as MCP servers and then used by any LLM client that supports the protocol~\cite{gan2025ragmcpmitigatingpromptbloat}. 
This design avoids repeated tool definitions, reduces maintenance overhead, and makes it easy to share tools across platforms. 
Its modular structure is suitable for CFD post-processing, which involves numerous specialized functions for tasks like data extraction, visualization, and performance analysis~\cite{krishnan2025advancingmultiagentsystemsmodel}. 
Through MCP, these varied functions can be unified under a single interface, offering a scalable and sustainable framework for LLM-driven CFD automation.

We introduce CFD-copilot, an automated framework designed to manage the CFD workflow from case setup to post-processing through natural language. 
This work refines and extends the LLM-driven foundation established in our prior research~\cite{dong2025autocfd} by proposing a dual-component architecture for enhanced automation and scalability. 
The first component is a domain-adapted LLM that acts as an intelligent interface to OpenFOAM, translating high-level user descriptions into executable simulation configurations. 
This relieves operators from needing to master the software's complex syntax. 
The second contribution is a highly extensible post-processing layer built on MCP. 
This MCP server decouples the LLM's reasoning from specific tool execution, creating a model-agnostic component that enables robust and scalable language-driven analysis. 
The resulting framework was evaluated on canonical aerodynamic benchmarks, including the NACA 0012 airfoil and the complex three-element 30P-30N high-lift configuration, to demonstrate its applicability to practical engineering problems.

This paper is organized as follows.
Section~\ref{sec:method} introduces the CFD-copilot framework, with a focus on its implementation for OpenFOAM-based workflows. 
Section~\ref{sec:results} presents experimental evaluations on two canonical aerodynamic benchmarks demonstrating the framework’s capabilities in automated simulation execution and post-processing tasks.
Section~\ref{sec:conclusions} summarizes our findings, discusses limitations, and outlines directions for future research in LLM-assisted CFD automation.

\section{Methodology}
\label{sec:method}

\subsection{CFD-copilot}

\begin{figure*}[ht]
    \centering
    \includegraphics[width=\textwidth]{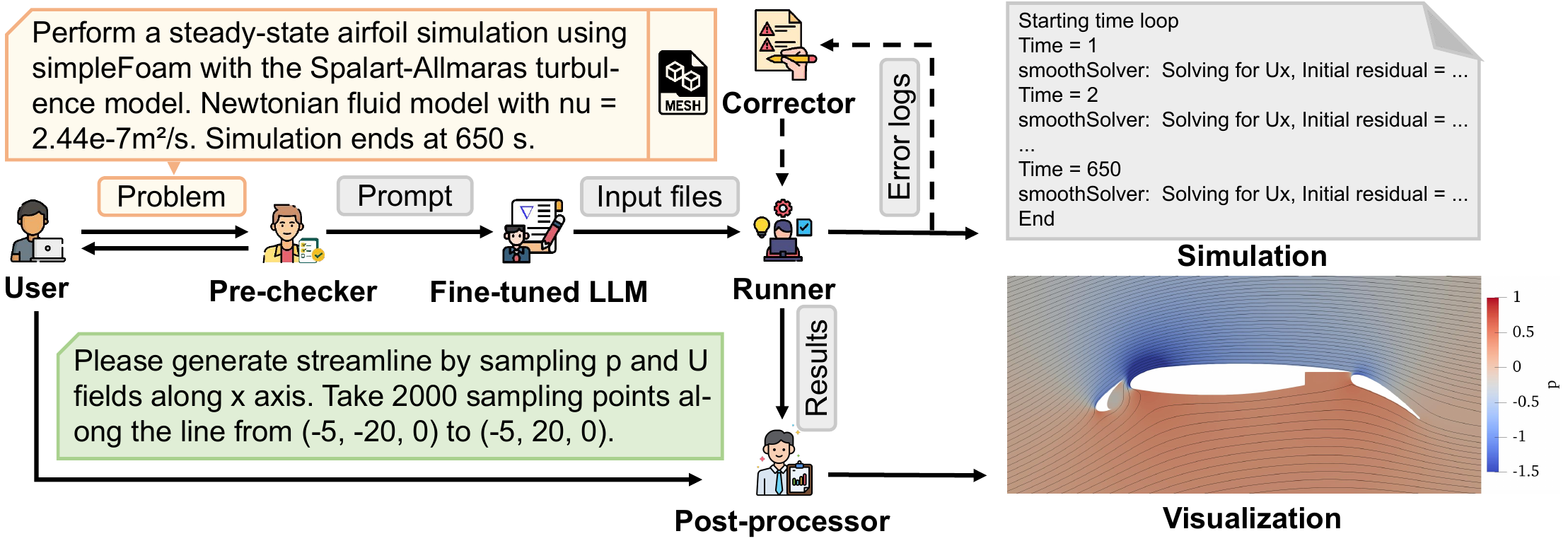}
    \caption{
    The framework of CFD-copilot. The workflow progresses from user input through a pre-checker, LLM-based generation of input files, simulation execution by the runner, and an iterative error correction loop involving the corrector. Upon successful simulation completion, the post-processor extracts and visualizes flow fields based on natural language instructions, enabling direct access to numerical results and high-fidelity visualizations.
    }
    \label{fig:multi-agent}
\end{figure*}

CFD-copilot is an integrated, multi-agent framework designed to automate the CFD workflow from simulation setup to post-processing. 
Building on our prior architecture for natural language-driven OpenFOAM execution~\cite{dong2025autocfd}, this work integrates a modular post-processing subsystem.
This extension enables a comprehensive automated workflow managed via natural language commands.
The overall framework, illustrated in Fig.~\ref{fig:multi-agent}, orchestrates specialized agents to manage the process from simulation configuration to data interpretation.

The initial workflow employs a self-correcting loop of four agents: \emph{pre-checker}, \emph{generator}, \emph{runner}, and \emph{corrector} to translate a user's natural language prompt into a viable simulation. 
The \emph{pre-checker} validates input and mesh files, enabling the LLM-powered \emph{generator} to produce OpenFOAM configurations. 
Subsequently, the \emph{runner} executes the simulation, while the \emph{corrector} analyzes error logs and iteratively prompts for adjustments until successful completion. 
This robust loop ensures a high success rate for the automated simulation setup.

To achieve domain-specific adaptation, we fine-tuned the Qwen3-8B model~\cite{yang2025qwen3technicalreport} using LoRA~\cite{hu2022lora} on the NL2FOAM dataset. 
This dataset contains 49205 pairs, where each pair maps a natural language description, mesh files, and input file templates to the corresponding OpenFOAM configurations.
The output is annotated with chain-of-thought~\cite{wei2022chain} reasoning to enhance generation quality and stability.
The resulting fine-tuned LLM serves as an interface between users and OpenFOAM, translating natural language specifications into executable CFD configurations without requiring users to master OpenFOAM's complex syntax and parameter structures. 

Following simulation, the framework transitions to the post-processing stage, which is orchestrated by the \emph{post-processor} agent. 
This \emph{post-processor} interprets natural language queries from the user, such as requests for data analysis or visualization. 
The agent leverages MCP, a model-agnostic architecture designed to decouple LLM reasoning from software execution. 
This approach enables a flexible and extensible framework for post-processing, as detailed in the following. 

\subsection{MCP-enabled post-processing}

\begin{figure*}[ht]
\centering
\includegraphics[width=\textwidth]{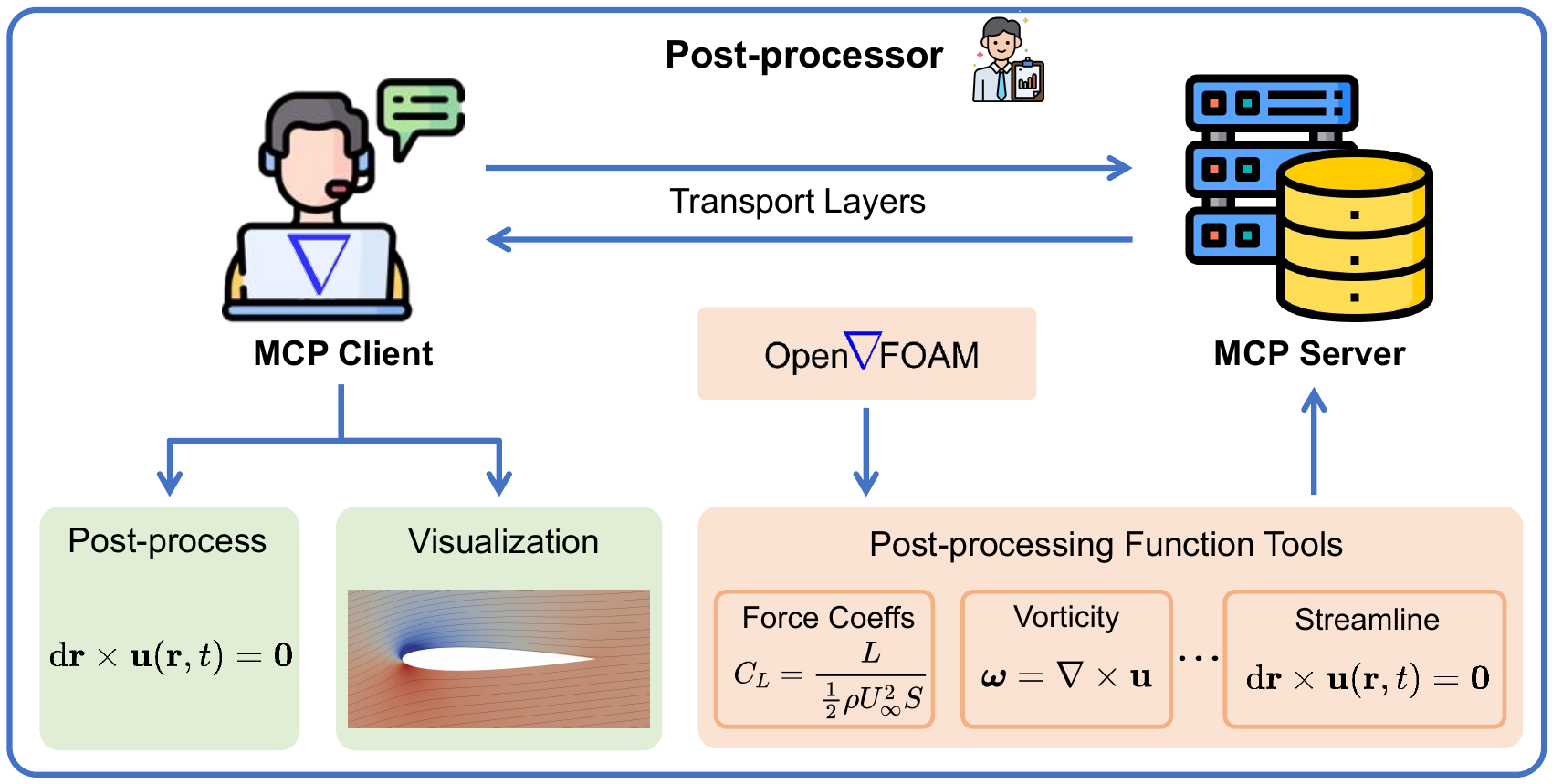}
\caption{Architecture of the MCP system in the \textit{Post-processor}. The MCP server hosts over 100 tool functions derived from the official OpenFOAM library, including utilities for calculating force coefficients, vorticity, and streamlines. The MCP client facilitates communication between the LLM and the server, enabling the execution of post-processing tasks and result visualization through natural language commands.
}
\label{fig:mcp-architecture}
\end{figure*}

The post-processing framework employs an MCP based client-server architecture to automate user interaction with OpenFOAM utilities, as illustrated in Fig.~\ref{fig:mcp-architecture}. 
The MCP server functions as a centralized repository, hosting over 100 validated and self-contained post-processing tools derived from OpenFOAM, such as utilities for calculating force coefficients (\texttt{forceCoeffs}) and generating vorticity fields (\texttt{vorticity}). 
On the other hand, the MCP client serves as an intelligent user interface. 
It integrates an LLM to interpret natural language commands, manage dialogue context, and orchestrate the post-processing workflow by transmitting structured requests to the MCP server.


The framework achieves scalability and modularity through the standardized design of tools hosted on the MCP server.
As exemplified in Fig.~\ref{fig:mcp-example}, each tool, such as \texttt{postProcess\_\allowbreak forceCoeffs}, comprises a structured annotation and execution logic. 
The annotation serves as both documentation and metadata, summarizing functionality and input arguments.
The client-side MCP engine leverages this metadata to identify the appropriate tool based on a user request.
Subsequently, the execution logic dynamically constructs and executes the required OpenFOAM command (e.g., \texttt{postProcess}) to ensure correct operation within the case environment.

\begin{figure}[!ht]
\centering
\begin{tcolorbox}[colback=blue!5!white, colframe=blue!0!black, title={An example tool in MCP server}]
\begin{singlespace}
\begin{lstlisting}[style=python_style]
@app.tool()
def postProcess_forceCoeffs(time: str, patches: str, ...) -> str:
    """
    Computes force and moment coefficients over a given
    list of patches, and optionally over given porous zones.
    Args:
        time(str): Specifies the range of time steps to process.
        patches(str): A list of patch names to sample.
        ...
    """
    func = "forceCoeffs"
    func_id = get_func_id(func, case_dir)
    function_content = f"""
        {func_id}
        {{
            type            forceCoeffs;
            libs            (forces);
            writeControl    writeTime;
            patches         ({patches});
            ...
        }}
    """
    write_function_objects(case_dir, function_content)
    solver = get_solver_name(case_dir)
    command = f"{solver} -postProcess -dict 
                system/postProcessingDict"
    command = add_latest_time_option(command, latestTime)
    command = add_time_option(command, time)
    return command
\end{lstlisting}
\end{singlespace}
\end{tcolorbox}
\caption{An example MCP server tool for computing force coefficients in OpenFOAM.}
\label{fig:mcp-example}
\end{figure}

The MCP client drives the workflow by translating user intent into executable actions. 
Upon receiving a natural language request, such as ``compute lift and drag coefficients on the wing surface at the last timestep,'' the client's LLM parses the query.
It then searches the metadata of all tools to select the appropriate utility, in this case, \texttt{forceCoeffs}. 
Following tool selection, the LLM maps the request details to the tool parameters, identifying ``wing surface'' for the patches argument and ``last timestep'' for the time argument.
Finally, the client transmits a structured MCP payload to the server, invoking the tool's execution logic to generate the result.

This client-server architecture and standardized tool protocol establish a highly modular and scalable system. 
Because each utility on the server is self-descriptive and independently callable via the MCP interface, the framework's capabilities can be easily expanded. 
New post-processing functions can be integrated by adding tool files to the server, requiring neither client modifications nor language model retraining. 
This separation of concerns ensures long-term maintainability and extensibility, enabling the framework to evolve alongside the underlying CFD software.

%


\section{Results}
\label{sec:results}

\subsection{Experimental setup}
\label{subsec:experimental_setup}

The framework's performance was evaluated on two canonical aerodynamic benchmarks: the 2D NACA 0012 airfoil~\cite{ladson1988effects} and the three-element 30P-30N high-lift configuration~\cite{klausmeyer1997comparative}. 
To ensure a valid assessment of the model's generalization capabilities, these cases were outside the fine-tuning dataset. 
We also provide results for cylinder wake case in the supplementary material.
The evaluation for each benchmark was conducted in two stages: automated simulation setup and subsequent automated post-processing. 
The \emph{generator} agent uses our fine-tuned Qwen3-8B model to produce OpenFOAM configurations. 
The remaining components are powered by the general-purpose Qwen3-32B~\cite{yang2025qwen3technicalreport} model.
For all experiments, we set the model temperature to 0.6, and reported metrics were averaged over 10 trials with a maximum of 10 correction iterations per trial. 
The multi-agent system was built based on MetaGPT v0.8.1~\cite{hong2024metagpt}, the post-processing layer utilized MCP v1.9.0~\cite{anthropic2025mcp}, and all simulations were executed using OpenFOAM-v2406~\cite{openfoamv2406opencfd}.

\subsection{NACA 0012}
\label{subsubsec:naca0012}

\begin{figure}[ht]
\centering
\includegraphics[width=0.5\columnwidth]{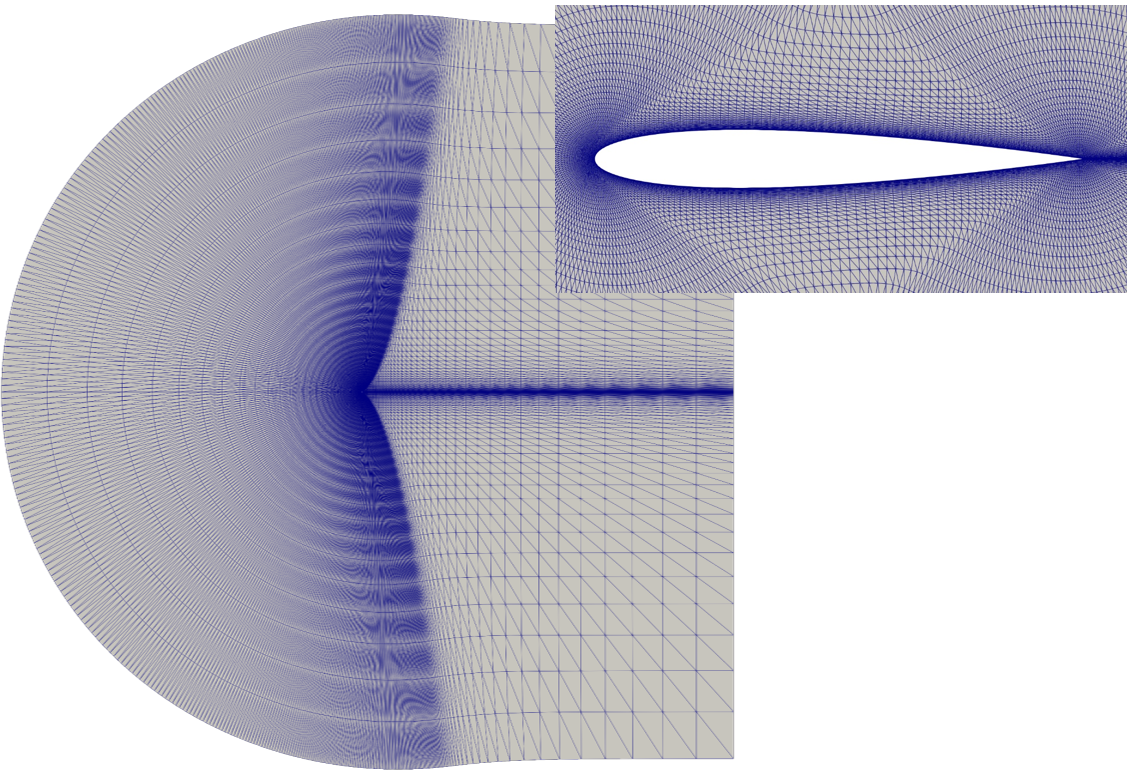}
\caption{
NACA~0012 airfoil mesh.
}
\label{fig:naca0012-mesh}
\end{figure}

The 2D NACA 0012 airfoil serves as a canonical benchmark for validating CFD configuring. 
The simulation parameters were set to match the experimental conditions~\cite{ladson1988effects}.
The freestream velocity is $U_\infty = 51.48~\mathrm{m/s}$. The fluid kinematic viscosity is $8.58 \times 10^{-6}~\mathrm{m^2/s}$.
This results in a Reynolds number of approximately $6 \times 10^6$ and a Mach number of 0.15. 
To reduce the influence of farfield boundary conditions on aerodynamic forces, the domain boundaries are placed approximately 500 chord lengths away from the airfoil.  
The computational domain extends from $x = -484.46$ to $x = 501$ and from $y = -507.81$ to $y = 507.81$, with a unit depth in the $z$ direction, giving a total size of about $985.46 \times 1015.62 \times 1$.  
The mesh consists entirely of hexahedral cells, forming a structured 2D mesh with 57344 cells. 
The geometry and computational mesh for this case are shown in Fig.~\ref{fig:naca0012-mesh}.

\subsubsection{Simulation setup and validation}

\begin{figure}[ht]
\centering
\includegraphics[width=0.5\columnwidth]{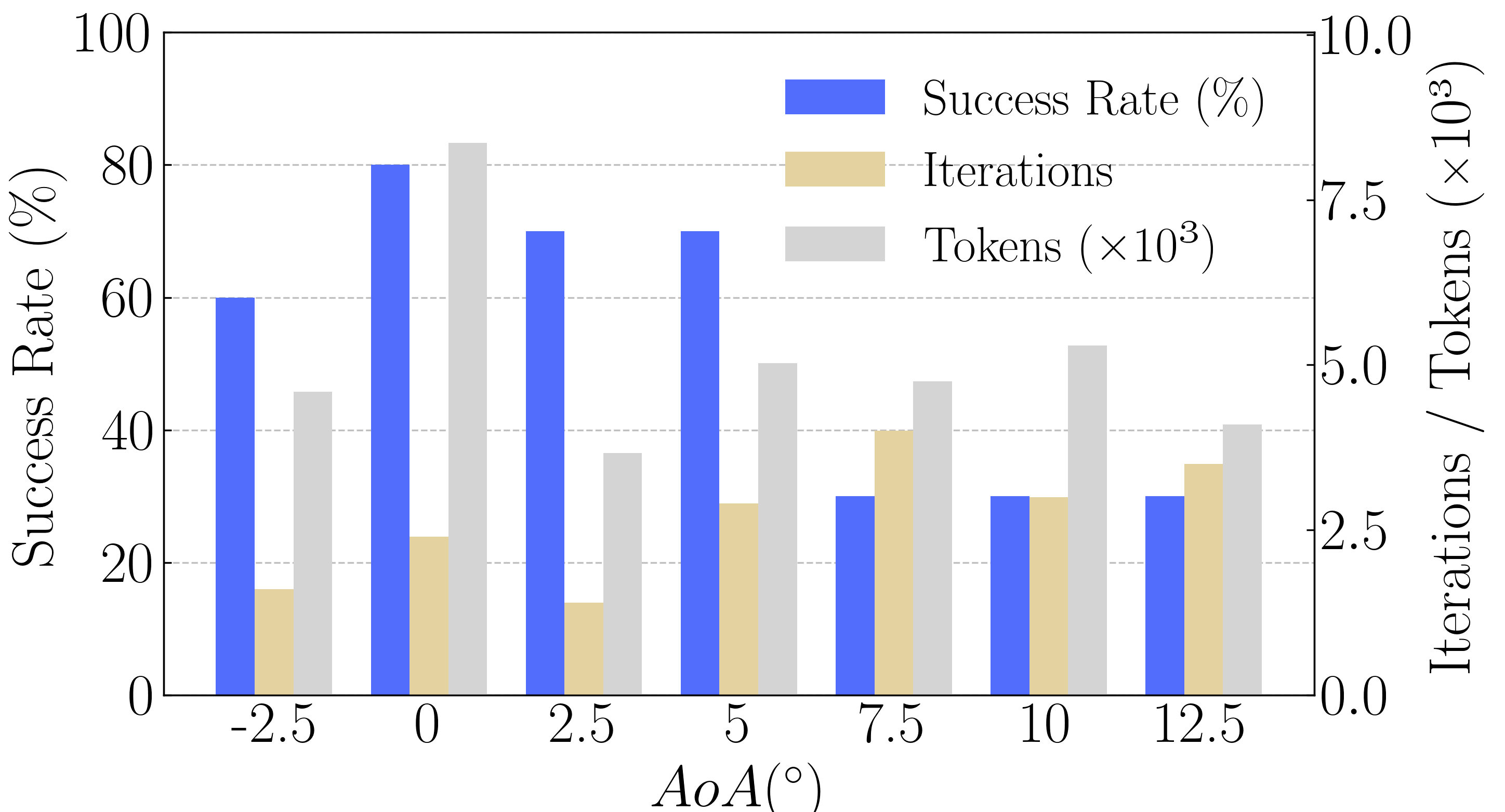}
\caption{Automation performance of the CFD-copilot framework on the NACA~0012 airfoil across different AoA. 
Bar chart shows success rate (\%), average correction iterations, and total token consumption (\( \times 10^3 \)).
}
\label{fig:naca0012-simulation-metrics}
\end{figure}

The initial objective was to test the framework's ability to generate and execute valid simulations. 
We tested on a range of angles of attack (AoA), from $-2.5^\circ$ to $12.5^\circ$. 
The automation performance, detailed in Fig.~\ref{fig:naca0012-simulation-metrics}, shows a high simulation success rate (S.R.) for AoA up to $5^\circ$. 
However, as AoA increases exceed $5^\circ$, the mesh lack the necessary resolution and structural adaptability to accurately resolve the evolving complex flow features, which contributes to the subsequent decline in simulation success rate.
As the flow becomes more complex at higher AoA, the success rate declines, and the average number of correction iterations increases, reflecting the agent's active error-handling as it adapts to challenging flow physics like separation. 
Token consumption remained modest throughout, indicating the process is computationally efficient.

The accuracy of the agent-generated simulations was evaluated against experimental data~\cite{ladson1988effects}.
The metrics presented in Tab.~\ref{tab:naca0012-simulation-metrics} include velocity $U$ and pressure $p$ field accuracies alongside the relative errors for lift $C_l$ and drag $C_d$ coefficients. 
For AoA up to \(10^\circ\), the velocity and pressure field accuracies remain above 95\%.  
The large relative error in the lift coefficient at 0$^\circ$ AoA is an expected artifact caused by dividing a small numerical deviation by a near-zero experimental lift value.
Overall, the simulation results are consistent with the experimental trend at low AoA. 
When AoA exceeds 10$^\circ$, errors in both lift and drag coefficients increase. 
This correlates with the onset of more complex flow physics, where the fixed computational mesh has insufficient adaptation to fully resolve the corresponding features. 
Overall, these results confirm the framework can autonomously configure simulations.

\begin{table}[htbp]
  \centering
  \caption{Flow field and aerodynamic accuracy of the CFD-copilot framework for the NACA~0012 airfoil across different AoA. 
  ``Accuracy'' measures solution reliability using the L2 norm $\epsilon$ between the LLM-based automated CFD solution and the benchmark, defined as $1-\epsilon$. 
  The reference values for the lift and drag coefficients used to compute the relative errors are taken from the experiments~\cite{ladson1988effects}. 
  }
  \label{tab:naca0012-simulation-metrics}
  \begin{tabular}{c S[table-format=2.2] S[table-format=2.2] S[table-format=2.2] S[table-format=2.2]}
    \toprule
    AoA & {$U$ accuracy (\%)} & {$p$ accuracy (\%)} & {Lift error (\%)} & {Drag error (\%)} \\
    \midrule
    -2.5° & 97.75 & 97.83 & 2.16  & 10.73 \\
    0°    & 97.56 & 98.53 & 97.57 & 6.01  \\
    2.5°  & 97.49 & 95.97 & 4.82  & 10.46 \\
    5°    & 98.55 & 99.09 & 2.49  & 12.77 \\
    7.5°  & 97.33 & 95.97 & 3.79  & 30.54 \\
    10°   & 96.37 & 93.20 & 6.64  & 32.21 \\
    12.5° & 89.82 & 71.95 & 28.11 & 37.29 \\
    \midrule
    Avg.  & 96.41 & 93.22 & 20.80 & 20.00 \\
    \bottomrule
  \end{tabular}
\end{table}

\subsubsection{Post-processing}

Following the successful generation of simulation data, we evaluated the framework's post-processing capabilities. 
The objective was to test the system's ability to translate high-level scientific questions into a concrete multi-step analysis workflow.
We selected the 10$^\circ$ AoA case as a representative example. 
This process is orchestrated by MCP, which interprets natural language prompts to first select and execute the appropriate OpenFOAM utility for data generation. 
Subsequently, the LLM processes user requests to generate Python scripts for visualization and further analysis of this data.

\begin{figure}[ht]
    \centering
    \begin{subfigure}[b]{0.5\columnwidth}
        \includegraphics[width=\textwidth]{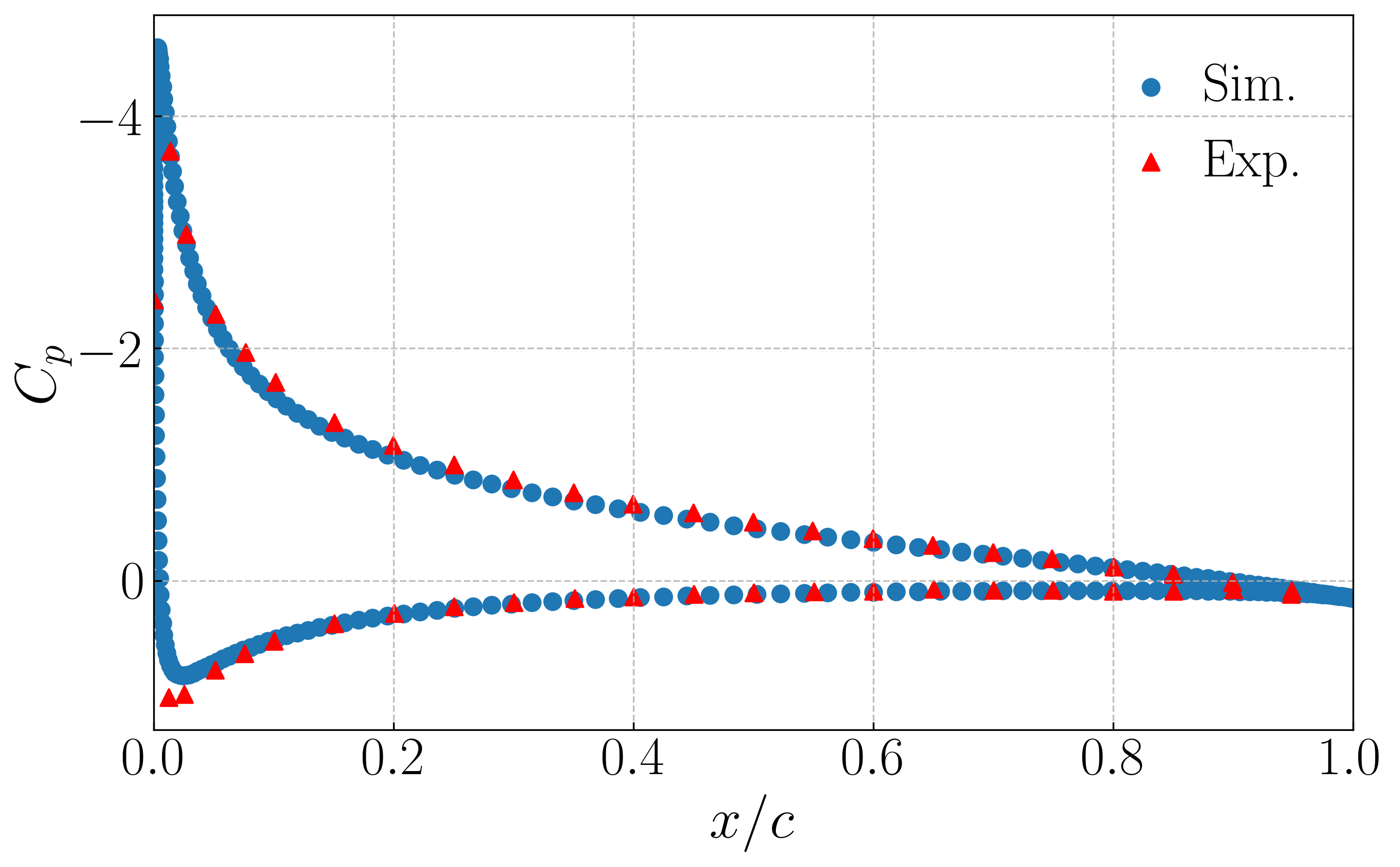}
        \caption{Pressure coefficient distribution along the chord of the NACA 0012 airfoil at $10^\circ$ AoA.}
        \label{fig:naca0012-cp-deg-10}
    \end{subfigure}
    \begin{subfigure}[b]{0.5\columnwidth}
        \includegraphics[width=\textwidth]{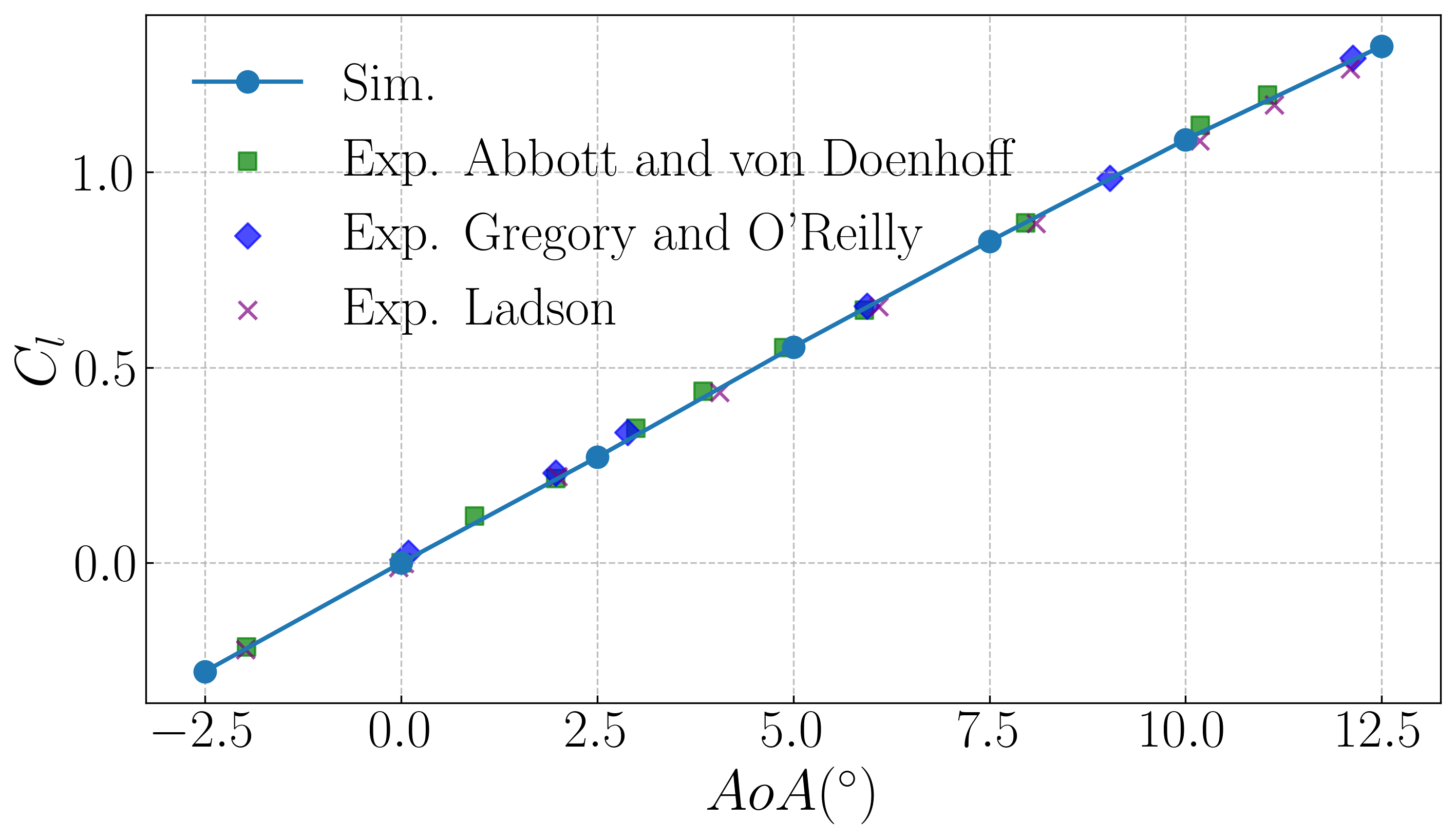}
        \caption{Lift coefficient $C_l$ versus AoA for the NACA 0012 airfoil.}
        \label{fig:naca0012-cl}
    \end{subfigure}
    \caption{Analysis of aerodynamic characteristics of the NACA 0012 airfoil, including pressure coefficient distribution and lift coefficient variation with AoA.}
    \label{fig:naca0012-analysis}
\end{figure}

The first task dealt with surface pressure distribution. 
The agent was guided by a two-prompt sequence. The initial prompt,
\begin{lstlisting}[style=text_style]
Please sample field p on the `walls' patch.
\end{lstlisting}
was interpreted by MCP to invoke the \texttt{postProcess\_surfaces\_sampledPatch} tool. 
This command extracted the pressure data along the airfoil's surface patch.
A second prompt, 
\begin{lstlisting}[style=text_style]
Please write a Python script to draw a scatter plot of normalized chord length and pressure coefficient.
\end{lstlisting}
instructed the LLM to generate a Python script. 
This script processed the raw data, calculated the non-dimensional pressure coefficient ($C_p$), and plotted the result shown in Fig.~\ref{fig:naca0012-cp-deg-10}. 
The resulting distribution shows the expected suction peak on the upper surface and matches well with reference data, confirming the framework's ability to perform targeted data extraction and visualization.

Building on this, the second task evaluated the computation of integrated aerodynamic forces. 
The prompt provided the necessary physical and geometric parameters:
\begin{lstlisting}[style=text_style]
Please compute force coefficients over `walls' patch at the latest time. Lift direction is (-0.1736 0.9848 0). Drag direction is (0.9848 0.1736 0). Pitch axis is (0 0 1). The magnitude of the free-stream velocity is 51.4815. Length of the wing is 1. Area of the wing is 1.
\end{lstlisting}
In response, MCP parsed the request and identified \texttt{postProcess\_forceCoeffs} as the correct utility. 
It then used the lift/drag vectors and reference values specified in the prompt to configure and execute the tool, calculating the lift and drag coefficients.
The resulting lift coefficients, plotted in Fig.~\ref{fig:naca0012-cl}, align with the experimental data~\cite{abbott2012theory,gregory1970low,ladson1988effects}, demonstrating the system's capacity to perform accurate quantitative analysis from a high-level command.

\begin{figure}[ht]
\centering
\includegraphics[width=0.5\columnwidth]{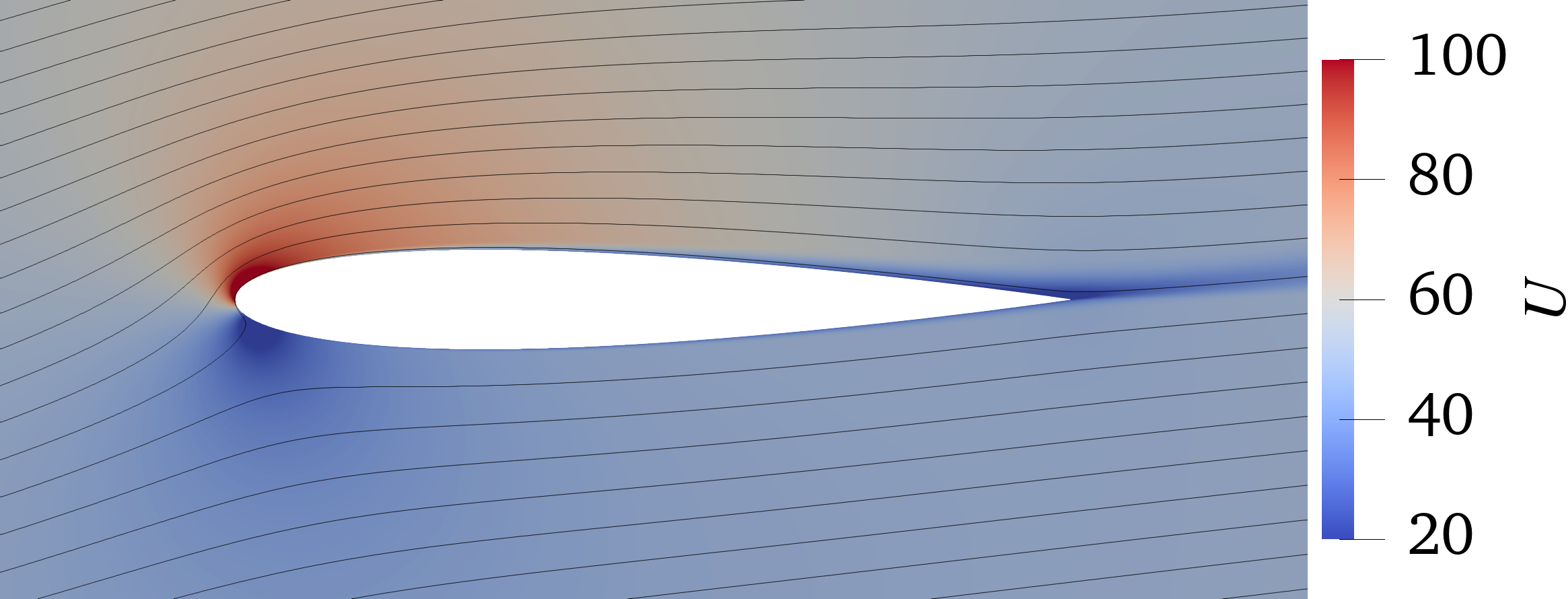}
\caption{Streamline visualization around the NACA 0012 airfoil at $10^\circ$ AoA.}
\label{fig:naca0012-streamline-U}
\end{figure}

The final task focused on qualitative flow field analysis using streamlines visualization.
This was initiated by the prompt,
\begin{lstlisting}[style=text_style]
Please generate streamline by sampling p and U fields along x axis. Take 1000 sampling points along the line from (-5, -20, 0) to (-5, 20, 0).
\end{lstlisting}
MCP interpreted this command and selected the \texttt{postProcess\_streamLine} tool.
It used the coordinates specified in the prompt to define the seeding line for integration of the velocity field.
The resulting visualization, shown in Fig.~\ref{fig:naca0012-streamline-U}, correctly depicts the flow accelerating over the suction surface and forming a wake downstream, providing qualitative confirmation of the underlying flow physics.

The NACA 0012 airfoil case demonstrates the flexibility and reliability of the MCP-enabled automated workflow. 
Across both simulation setup and post-processing, the system successfully translates high-level natural language instructions into precise, multi-step CFD operations. 
It autonomously orchestrates a chain of discrete actions, from invoking specific OpenFOAM utilities for data extraction and quantitative calculation to generating custom python scripts for visualization. 
Crucially, these tasks were completed without requiring the user to handle low-level software commands, highlighting the system’s potential to significantly simplify and accelerate CFD workflows.

\subsection{Three-Element 30P-30N Airfoil}
\label{subsubsec:30p30n}

\begin{figure}[ht]
\centering
\includegraphics[width=0.5\textwidth]{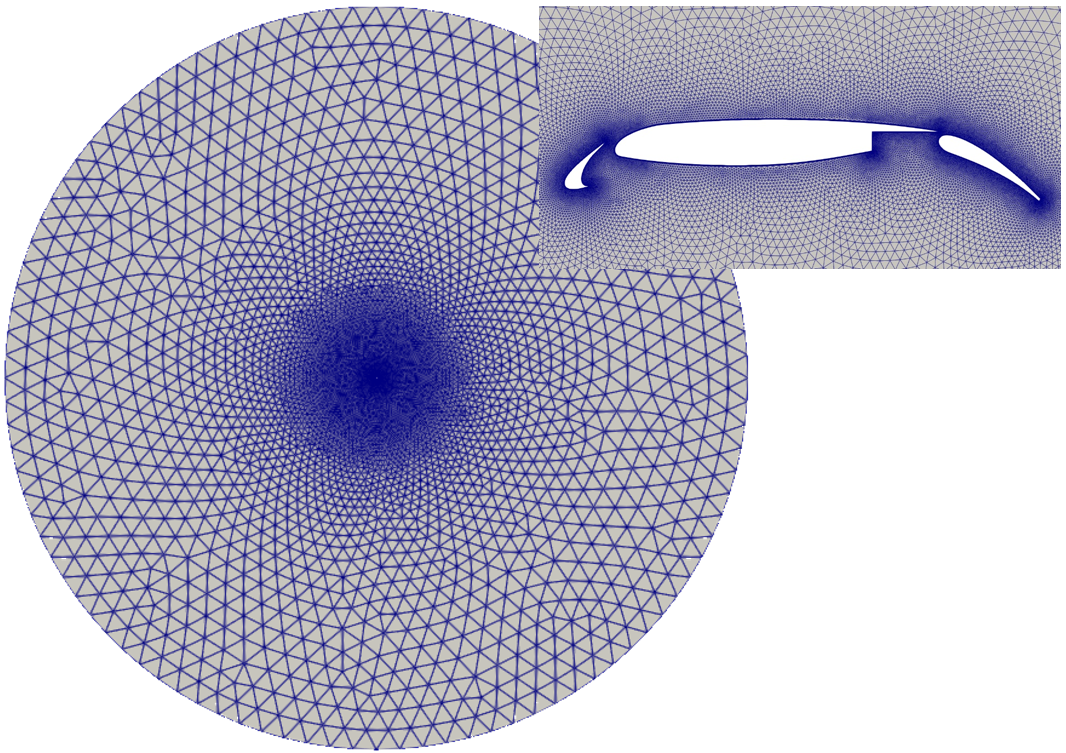}
\caption{Computational mesh for the three-element 30P-30N airfoil.}
\label{fig:30p30n-mesh}
\end{figure}

The three-element 30P-30N airfoil is a widely studied high-lift configuration consisting of a slat, main element, and flap. 
Owing to its complex flow features, such as strong boundary layer interactions, flow separation, and wake merging, this geometry has been extensively investigated~\cite{pascioni2014experimental}.
The availability of high-fidelity reference data makes the 30P-30N a standard benchmark for assessing the accuracy and robustness of CFD methods.
The computational domain is a circle with a radius of 150 chord lengths, which is discretized using a hybrid mesh of approximately 100,300 cells, as shown in Fig.~\ref{fig:30p30n-mesh}.

\subsubsection{Simulation setup and validation}

The primary challenge was to autonomously generate a valid solver configuration for the 30P-30N airfoil at an 8.12$^\circ$ AoA.
Our fine-tuned model achieved an 80\% simulation completion rate (C.R.) and, after iterative corrections, a 10\% final success rate (S.R.), as detailed in Tab.~\ref{tab:30p30n-automation}.
Although the success rate is low, it is a noteworthy achievement for a fully automated process. 
To contextualize this performance, we conducted a direct comparative benchmark by replacing our model with two much larger, general-purpose LLMs, Qwen3-Next-80B and Qwen3-235B~\cite{yang2025qwen3technicalreport}. 
Under identical conditions, these state-of-the-art models only achieved a 10\% completion rate and failed to produce a single converged solution, resulting in a 0\% success rate.

\begin{table}[ht]
\centering
\caption{Automation performance metrics for the three-element 30P-30N airfoil at $8.12^\circ$ AoA.}
\label{tab:30p30n-automation}
\begin{tabular}{c|cccc|cccc}
\toprule
AoA & C.R. & S.R. & Iters. & Tokens & U acc. & P acc. & $C_l$ err. & $C_d$ err. \\
\midrule
$8.12^\circ$ & 80\% & 10\% & 4.8 & 4536 & 93.14\% & 88.44\% & 2.58\% & 16.42\% \\
\bottomrule
\end{tabular}
\end{table}

The performance gap is primarily due to the sophistication of the generated solver configurations. 
The general-purpose LLMs tended to produce overly simplistic setups, such as defaulting to a basic ``Gauss linear'' scheme for all discretizations, which is inadequate for this case. 
In contrast, our fine-tuned model demonstrated its domain knowledge by applying variable-specific schemes. 
For instance, it assigned ``cellLimited Gauss linear 1'' to velocity and turbulence kinetic energy, enhancing numerical stability.

For the successful run, the solution is accurate.
The lift coefficient, $C_l$, exhibited a relative error of 2.58\% compared to the experimental data~\cite{klausmeyer1997comparative}.
The larger drag coefficient error, 16.42\%, is expected.
This discrepancy arises because the framework, which focuses on automating solver configuration rather than mesh adaptation, relies on a fixed, moderately-resolved mesh. 
Although this mesh is sufficient for accurate lift prediction, it lacks the fine resolution required to fully resolve the drag-sensitive wake regions.

\subsubsection{Post-processing}

The next stage tested the framework's ability to perform analysis across the multiple distinct surfaces of the 30P-30N airfoil. 
The challenge was to guide the post-processing of the successfully converged simulation data using simple natural language prompts, requiring the system to manage and aggregate information from the slat, main element, and flap components. 
This again leveraged MCP to translate user intent into a sequence of tool invocations and script generation.

\begin{figure}[!ht]
    \centering
    \begin{subfigure}[b]{0.5\columnwidth}
        \centering
        \includegraphics[width=\textwidth]{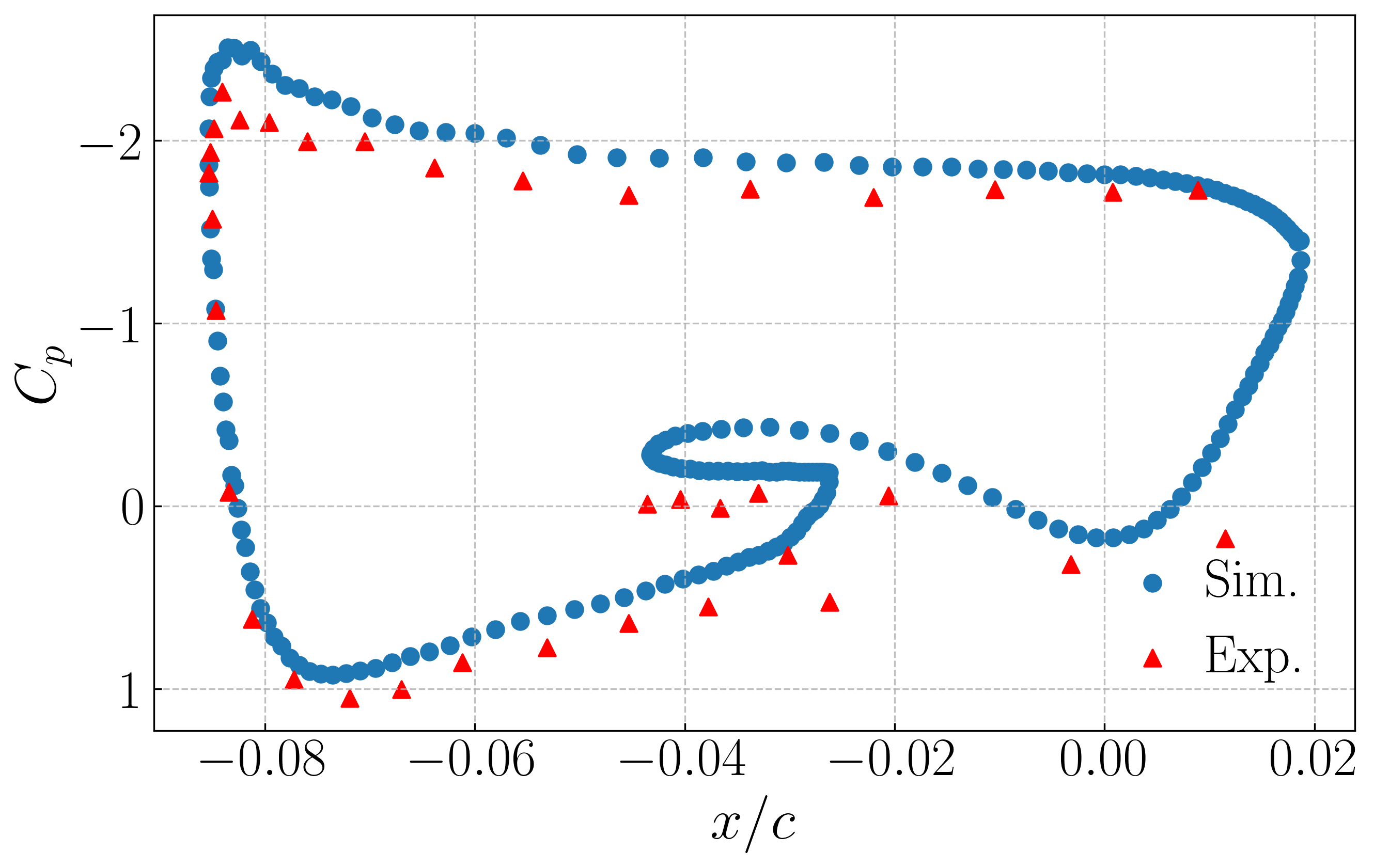}
        \caption{Pressure coefficient distribution along the slat surface.}
        \label{fig:30p30n-slat-cp}
    \end{subfigure}
    \begin{subfigure}[b]{0.5\columnwidth}
        \centering
        \includegraphics[width=\textwidth]{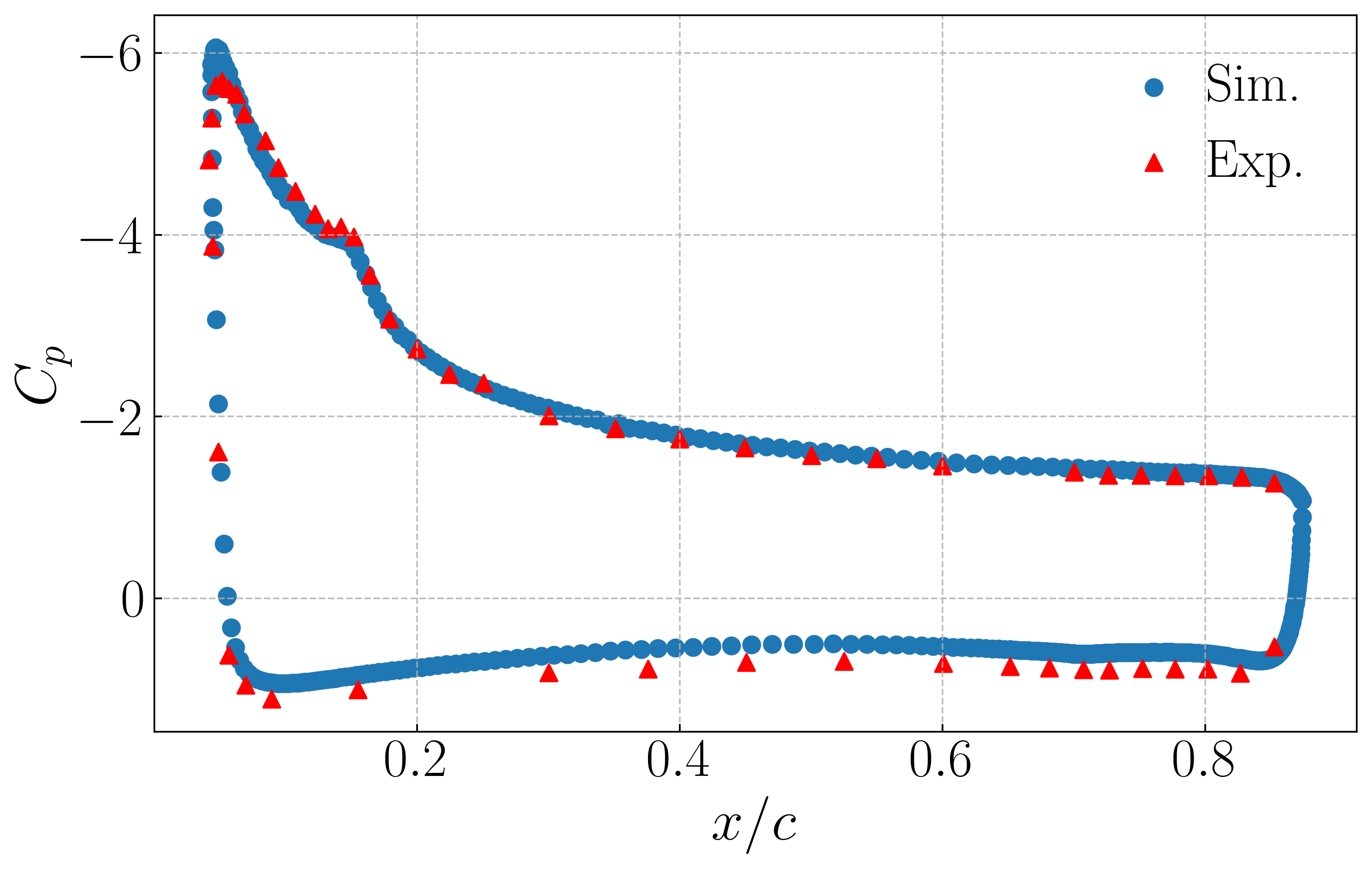}
        \caption{Pressure coefficient distribution along the slat surface.}
        \label{fig:30p30n-main-cp}
    \end{subfigure}
    \begin{subfigure}[b]{0.5\columnwidth}
        \centering
        \includegraphics[width=\textwidth]{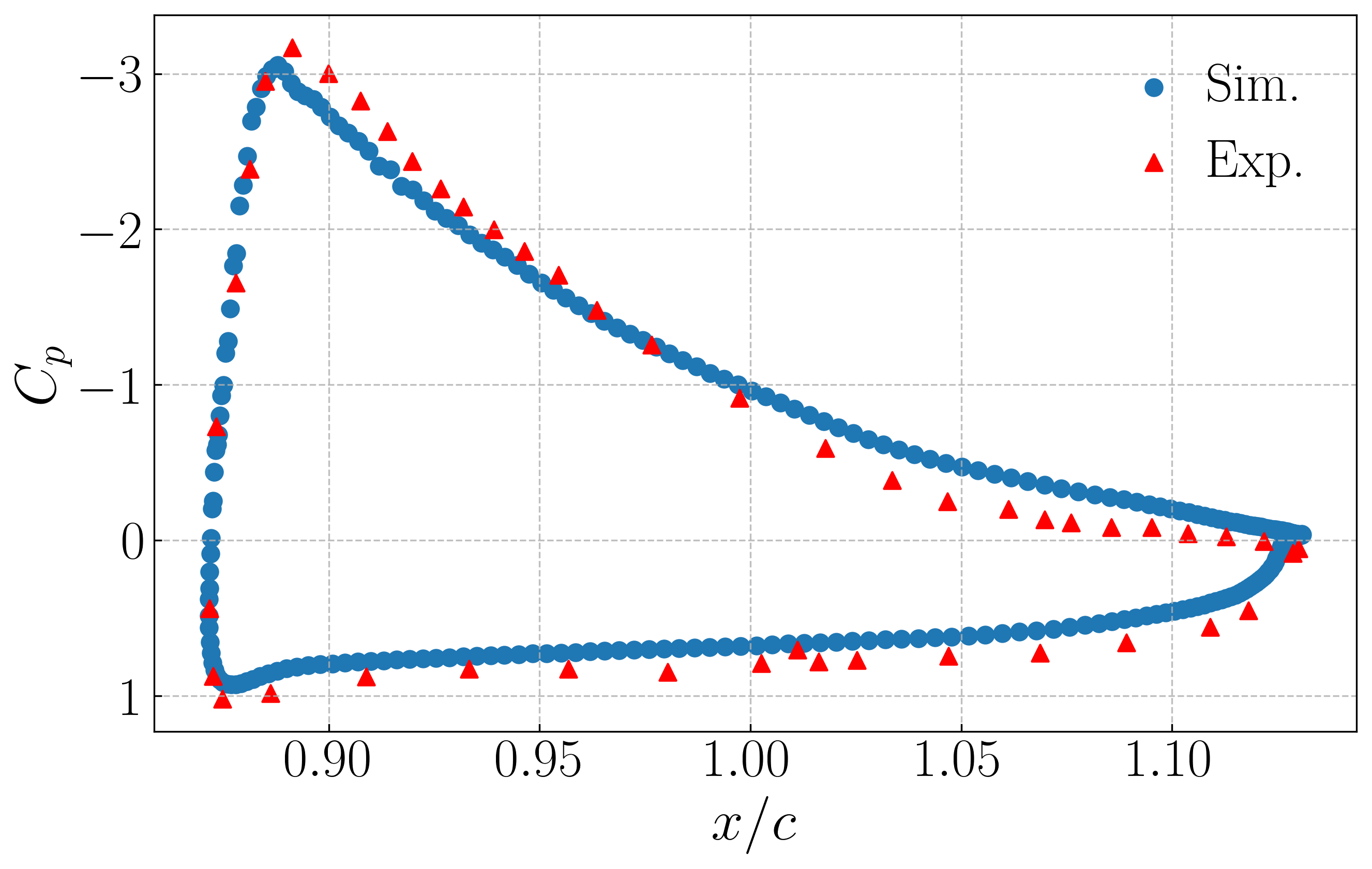}
        \caption{Pressure coefficient distribution along the flap surface.}
        \label{fig:30p30n-flap-cp}
    \end{subfigure}
    \caption{Pressure coefficient distribution along each surface of the three-element 30P-30N airfoil at $8.12^\circ$ AoA.}
    \label{fig:30p30n-cp}
\end{figure}

First, the agent was tasked to yield the pressure distribution on each of the three elements individually. 
This was initiated with prompts specifying each patch,
\begin{lstlisting}[style=text_style]
Please sample field p on the `wall_slat'(or `wall_airfoil', `wall_flap') patches.
\end{lstlisting}
For each prompt, MCP correctly invoked the \texttt{postProcess\_surfaces\_sampledPatch} tool, generating a separate raw data file for the specified surface. 
After extracting data for all three components, a single, general command, 
\begin{lstlisting}[style=text_style]
Please write a Python script to draw a scatter plot of normalized chord length and pressure coefficient.
\end{lstlisting}
instructed the LLM to generate a script that automatically located, processed, and plotted the data from all three files. 
The resulting distributions in Fig.~\ref{fig:30p30n-main-cp} illustrate the characteristic features of a high-lift system, including the strong suction peak on the slat and the pressure recovery on the main element, with results matching well with experiments~\cite{klausmeyer1997comparative}. 
This demonstrates the ability to manage and visualize data from multiple geometric components in a unified workflow.

\begin{figure}[!ht]
    \centering
    \begin{subfigure}[b]{0.5\columnwidth}
        \includegraphics[width=\textwidth]{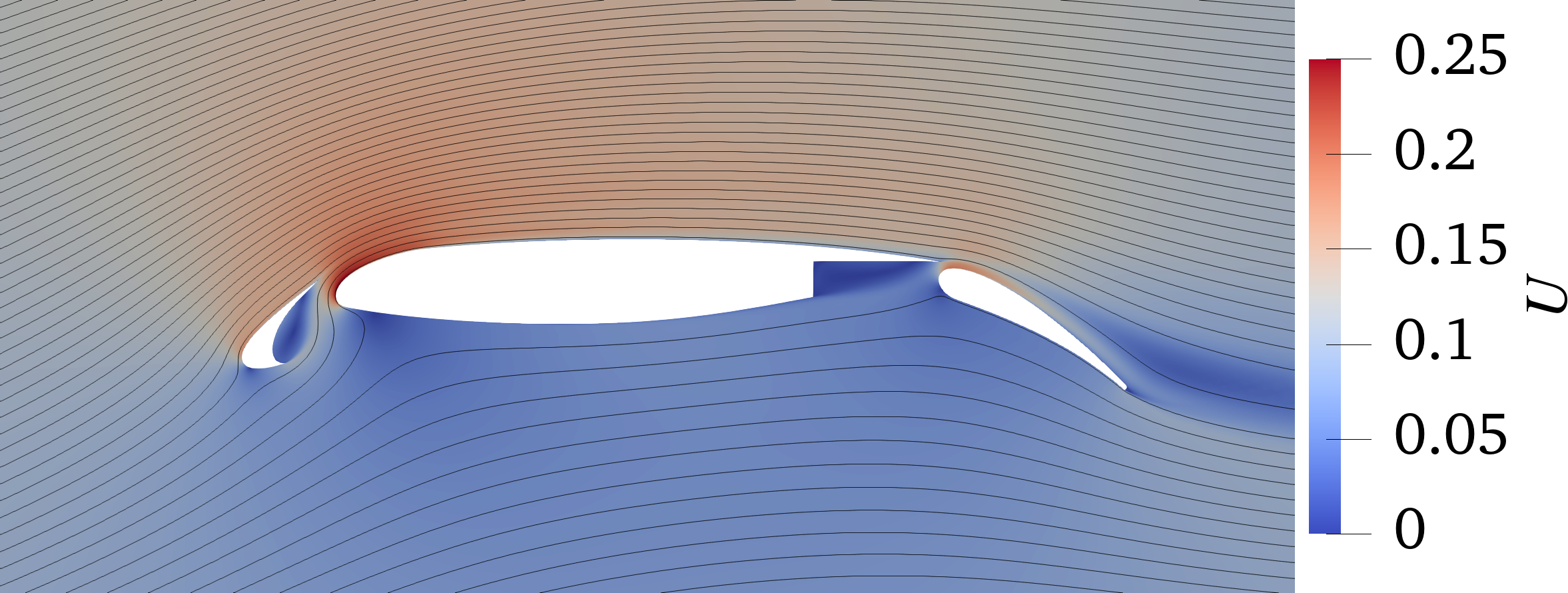}
        \caption{}
        \label{fig:30p30n-streamLine-U}
    \end{subfigure}
    \begin{subfigure}[b]{0.5\columnwidth}
        \includegraphics[width=\textwidth]{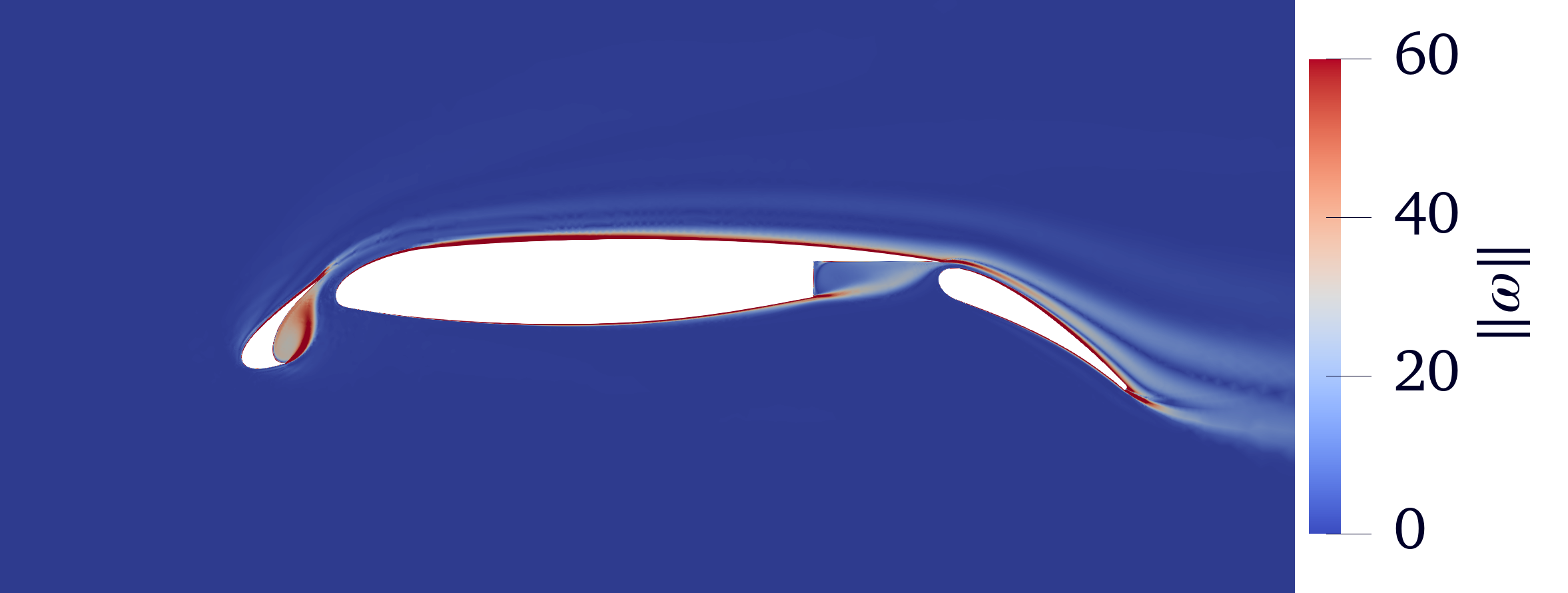}
        \caption{}
        \label{fig:30p30n-vorticity}
    \end{subfigure}
    \caption{
    (a) Streamline visualization around the three-element 30P-30N airfoil, showing flow acceleration over the slat, circulation around the main element, and wake formation behind the flap. 
    (b) Instantaneous vorticity field around the three-element 30P-30N airfoil, highlighting vortical structures near the slat cove, shear layer roll-up in the flap wake, and wake merging downstream.
    }
    \label{fig:30p30n-visualization}
\end{figure}

Qualitative analysis of the complex flow field was requested through streamline and vorticity visualizations via prompts
\begin{lstlisting}[style=text_style]
Please generate streamline by sampling p and U fields along x axis. Take 2000 sampling points along the line from (-5, -20, 0) to (-5, 20, 0).
\end{lstlisting}
and
\begin{lstlisting}[style=text_style]
Please calculate the vorticity of the airfoil.
\end{lstlisting}
In response, MCP invoked the appropriate OpenFOAM utilities, \texttt{postProcess\_streamLine} and \texttt{postProcess -func vorticity}, respectively. 
The agent configured these tools and executed them to generate the necessary field data for visualization. 
The resulting images in Fig.~\ref{fig:30p30n-visualization} display the intricate flow structures, such as acceleration through the slat-main element gap and wake merging downstream, which align with established high-lift aerodynamic principles. 

The 30P-30N airfoil case further validates the MCP-enabled workflow. 
First, it highlights the advantage of a domain-specific fine-tuned model, which succeeded in generating a valid solver configuration where much larger, general-purpose models failed. 
Second, it demonstrates that the prompt-driven MCP workflow scales effectively to complex, multi-body geometries. 
The framework seamlessly managed multi-surface data extraction, aggregation, and advanced visualizations, enabling a sophisticated analysis campaign without exposing the user to the underlying operational complexity.

\section{Conclusions}
\label{sec:conclusions}

We presented CFD-copilot, a natural language-driven framework for end-to-end automation of CFD simulations. 
The implementation for the initial case setup is managed by a multi-agent, self-correcting system organized around a central generator agent. 
This agent is powered by a domain-adapted Qwen3-8B model, fine-tuned to translate high-level prompts into valid OpenFOAM configurations. 
For the subsequent post-processing phase, the framework employs a modular client-server architecture built on MCP, which allows an LLM-driven client to orchestrate a server hosting a comprehensive suite of independent OpenFOAM utilities.
The performance of this integrated, language-driven system was evaluated on two canonical aerodynamic benchmarks, the 2D NACA 0012 airfoil and the three-element 30P-30N airfoil. 

Our validations confirms the efficacy of the proposed workflow. 
For the NACA 0012 airfoil case, the system achieved an average success rate of 52.86\%, with velocity and pressure accuracies of 96.41\% and 93.22\%, respectively. 
For the more complex multi-element 30P-30N airfoil, the system demonstrated scalability to realistic high-lift configurations. 
It achieved a 10\% success rate, with velocity and pressure accuracies of 93.14\% and 88.44\%.
In direct contrast, larger, general-purpose models failed to produce a single converged solution, highlighting that domain adaptation is critical for handling complex simulations.

Furthermore, leveraging the MCP architecture, the framework demonstrated a scalable and robust post-processing capability across both benchmarks.
It successfully translated abstract user prompts into a concrete series of tool-specific commands to perform tasks ranging from surface pressure plotting and multi-patch force coefficient aggregation to streamline and vorticity visualization. 
This confirms that the MCP's decoupling of reasoning from execution provides a viable path for managing complex, multi-step analytical workflows.

This work establishes a robust proof-of-concept for a complete, language-driven simulation cycle. 
By reducing reliance on domain-specific syntax and manual setup, CFD-copilot lowers the barrier to entry and allows operators to focus on the scientific objectives of their analysis rather than the intricacies of implementation. 
This approach has the potential to accelerate the design and analysis process in engineering.

Despite these successes, a primary limitation remains the reliability of generating robust solver configurations for highly complex flows, as indicated by the modest success rate for the 30P-30N case. 
Future work will focus on addressing this challenge by exploring advanced strategies, such as reinforcement learning, to refine solver parameters. 
The goal is to enhance the physical consistency and numerical stability of the generated setups, thereby improving success rates for a broader range of demanding engineering simulations.

\section*{Acknowledgments}

This work has been supported in part by the National Natural Science Foundation of China (Nos.~52306126, 12525201, 12432010, 12588201), and the National Key R\&D Program of China (Grant No.~2023YFB4502600). 



\bibliographystyle{elsarticle-num}
\biboptions{sort&compress}
\bibliography{CFD-copilot}

\end{document}